\RequirePackage{fix-cm}
\documentclass[twocolumn,final]{svjour3}           
\smartqed  
\usepackage{graphicx}
\usepackage{mathptmx}      
\usepackage[utf8]{inputenc}
\usepackage{amsmath}			
\usepackage{amssymb}			
\usepackage{graphicx}			
\usepackage[dvipsnames]{xcolor}				
\usepackage{physics}            
\usepackage{siunitx}			
\usepackage{tikz}
\usetikzlibrary{calc, decorations.markings, decorations.pathmorphing, decorations.pathreplacing, decorations.fractals, arrows.meta,calendar,intersections,mindmap,folding,patterns,plothandlers,plotmarks,shapes,shadings,shadows,topaths,trees}
\usepackage{pgf}
\usepackage{ulem}
\pagecolor{white}
\usepackage{comment}
\usepackage{textcomp}           
\usepackage[numbers,square,comma,sort&compress,nonamebreak]{natbib}
\newcommand{\ie}{i.\,e.}
\newcommand{\eg}{e.\,g.}
\newcommand{\eqspace}{\:}	
\newcommand{\sref}[1]{section~\ref{#1}}

\newcommand{\fref}[1]{figure~\ref{#1}}

\definecolor{sicolor}{rgb}{0,0,0}

\newcommand{\defeq}{\mathrel{\vcenter{%
			\baselineskip0.5ex\lineskiplimit0pt\hbox{%
				\scriptsize.}\hbox{\scriptsize.}}} =}					
\newcommand{\half}{\tfrac{1}{2}}							
\newcommand{\velocity}{u}									
\newcommand{\dynvisc}{\eta}									
\newcommand{\shearrate}{\dot{\gamma}}						
\newcommand{\capillarynumber}{\mathrm{Ca}}                  
\newcommand{\shearmodulus}{\mu}                             
\newcommand{\poissonratio}{\nu}                             
\newcommand{\youngsmodulus}{E}                              
\newcommand{\taylordeformation}{D}                          
\newcommand{\channelradius}{R_\mathrm{ch}}					
\newcommand{\cellradius}{R}									
\newcommand{\shellindex}{_\mathrm{c}}
\newcommand{\coreindex}{_\mathrm{n}}
\newcommand{\effindex}{_\mathrm{eff}}
\newcommand{\cellindex}{}
\newcommand{\radiuscell}{\cellradius\shellindex}
\newcommand{\coreradius}{R\coreindex}						
\newcommand{\radiuscore}{\coreradius}
\newcommand{\youngsshell}{\youngsmodulus\shellindex}		
\newcommand{\youngscore}{\youngsmodulus\coreindex}			
\newcommand{\youngseffective}{\youngsmodulus\effindex}		
\newcommand{\shearmodshell}{\shearmodulus\shellindex}
\newcommand{\shearmodcore}{\shearmodulus\coreindex}
\newcommand{\shearmodeff}{\shearmodulus\effindex}
\newcommand{\volumeshell}{V\shellindex}						
\newcommand{\volumecore}{V\coreindex}						
\newcommand{\volumecell}{V\cellindex}						
\newcommand{\stiffnessratio}{\gamma}						
\newcommand{\sizeratio}{\lambda}							
\newcommand{\offsetratio}{d}								
\newcommand{\deformation}{\delta}							
\newcommand{\force}{F}										
\newcommand{\strain}{\epsilon}								

\usepackage{hyperref}
\hypersetup{
	pdftitle={A mechanically homogeneous equivalent of real cells},    
	pdfauthor={S. Wohlrab and S. J. M\"uller and S. Gekle},     
	colorlinks=true,		
	linkcolor=blue,			
	citecolor=blue,			
	filecolor=magenta,		
	urlcolor=cyan			
}
\begin{document}
\title{Mechanical complexity of living cells can be mapped onto simple homogeneous equivalents}
%
%
%
%
%
\author{Sebastian Wohlrab$^\star$
		\and Sebastian J. M\"uller$^\star$
	    \and Stephan Gekle
}
%
%
\institute{Sebastian Wohlrab$^\star$ \and Sebastian J. M\"uller$^\star$ \and Stephan Gekle
		   \at Theoretical Physics VI, Biofluid Simulation and Modeling, University of Bayreuth, 95440 Bayreuth, Germany \\
           \email{stephan.gekle@uni-bayreuth.de} \\
           $^\star$: Sebastian Wohlrab and Sebastian J. M\"uller contributed equally.
}
\date{ }
%
%
\maketitle
\begin{abstract}
	Biological cells are built up from many different constituents of varying size and stiffness which all contribute to the cell's mechanical properties.
	Despite this heterogeneity, in the analysis of experimental measurements such as atomic force microscopy or microfluidic characterisation a strongly simplified homogeneous cell is typically assumed and a single elastic modulus is assigned to the entire cell.
	This ad-hoc simplification has so far mostly been used without proper justification.
	Here, we use computer simulations to show that indeed a heterogeneous cell can effectively be replaced by a homogeneous equivalent cell with a volume averaged elastic modulus.
	To study the validity of this approach, we investigate a hyperelastic cell with a heterogeneous interior under compression as well as in shear and channel flow, mimicking atomic force and microfluidic measurements, respectively.
	We find that the homogeneous equivalent cell reproduces quantitatively the behavior of its inhomogeneous counterpart, and that this equality is largely independent of the stiffness or spatial distribution of the heterogeneity.
\end{abstract}
\section{Introduction}
\label{sec:introduction}
Despite their internal complexity, the mechanics of biological cells is often approximated as a homogeneous elastic body either when analyzing experimental data or during finely resolved computer simulations.
One of the main techniques for characterizing the mechanical properties of cells is atomic force microscopy \cite{fischer-friedrich_quantification_2015,guz_if_2014,lulevich_cell_2006,lulevich_deformation_2003,ladjal_atomic_2009,kiss_elasticity_2011,hecht_imaging_2015,sancho_new_2017,muller_hyperelastic_2021}.
Other micromechanical evaluation techniques include the flow through highly confined microchannels~\cite{urbanska_comparison_2020,otto_real-time_2015,fregin_high-throughput_2019,rowat_nuclear_2013,lange_unbiased_2017,lange_microconstriction_2015}, or mechanical testing in larger channels~\cite{gerum_viscoelastic_2022-2}.
Both kinds of experiments are most commonly analyzed using a mechanical model which treats the entire cell as one continuous entity endowed with a single elastic modulus. 
This simple cell model has also been used in a series of computer simulations \cite{rosti_rheology_2018,saadat_immersed-finite-element_2018, muller_hyperelastic_2021}.

At the same time, however, it is known that the different constituents of the cell, \eg, the cortex, membrane, and nucleus, all have different mechanical properties~\cite{cordes_prestress_2020-1,zhelev_role_1994,lange_unbiased_2017,lykov_probing_2017-1, mietke_extracting_2015-1, caille_contribution_2002, cao_evaluating_2013}.
It is thus tempting to ask why the simplistic assumption of a homogeneous cell appears to work so surpisingly well in many situations.

In this work, we therefore systematically probe the possibility to substitute any inhomogeneously constituted cell with a simple homogeneous cell with an effective elasticity.
For that, we first construct a well-defined inhomogeneous cell with an inclusion, \eg, a nucleus, of variable stiffness (Young's modulus or shear modulus), size, and position.
In addition, we build an inhomogeneous cell with a spatially random stiffness distribution.
From the volume averaged mean of the constituents' Young's moduli we define an effective Young's modulus of a homogeneous equivalent cell.
With these at hand, we perform AFM compression simulations as well as microfluidic shear flow and pipe flow computations.
We find excellent agreement of the resulting force versus deformation behavior in compression and the strain versus fluid forces behavior in flow.
Through variation of stiffness, size, position, and shape, of the inhomogeneity we show that neither of these factors have a significant impact on the cell's mechanical behavior.
Any kind of intracellular mechanical diversity can hence be effectively described using our proposed homogeneous equivalent cell.

\section{Methods and setup}
\label{sec-methods}

\subsection{Inhomogeneous cell with nucleus}
\label{sec-methods-nucleus-cell-model}
\label{sec-methods-inhomogeneous-cell-model}
As model for a well-defined inhomogeneous cell, we use a cell with a stiffer nucleus inside.
We model the nucleate cell as a sphere of radius $\cellradius$ which contains a spherical inclusion of radius $\radiuscore$ inside the cell volume, as shown in \fref{fig-cell-model}(a).
It is labeled ``Nucleus'' in the plot.
We tetrahedralize both volumes and apply the neo-Hookean strain energy computations from \cite{muller_hyperelastic_2021} in both parts. (We achieved similar results using Mooney-Rivlin strain energy computations, see Appendix 1)
Properties of the whole cell are denoted without subscript, properties of the nucleus and the cytoskeleton by the subscripts ``$\mathrm{n}$'' and ``$\mathrm{c}$'', respectively.
The Poisson's ratio is $\poissonratio=0.48$ in all simulations, which ensures sufficient incompressibility while maintaining numerical stability.
To parametrize the stiffness we choose the Young's moduli $\youngscore$ and $\youngsshell$ of the inhomogeneity and the shell, respectively.

For our systematic analysis, we further define the stiffness ratio and the size ratio
\begin{align}
	\label{eq-def-stiffness-ratio}
	\stiffnessratio=\frac{\youngscore}{\youngsshell} \quad \mathrm{and} \quad \sizeratio = \frac{\radiuscore}{\radiuscell} \eqspace ,
\end{align}
with $\stiffnessratio>1$ describing an inhomogeneity stiffer than the rest of the cell and $0<\sizeratio<1$.
An additional offset $\offsetratio$ of the inhomogeneity from the cell's geometrical center is given in units of the cell radius.
Through variation of the control parameters $\stiffnessratio$, $\sizeratio$, and $\offsetratio$, any kind of spherical inclusion into the cell volume is covered.
We discuss the effect of an ellipsoidal inhomogeneity in the last paragraph of \sref{sec-results-compression}.
\\
As a reference configuration, from which variations of the control parameters start, we choose $\stiffnessratio=2$, $\sizeratio=\frac{1}{2}$, and $\offsetratio=0$.

\subsection{Random inhomogeneous cell model}
\label{sec-methods-random-cell-model}
In addition to the well-defined inhomogeneous system of \sref{sec-methods-nucleus-cell-model}, we create a random inhomogeneous cell by randomly assigning a stiffness ratio $\stiffnessratio_i \in \qty[1,10]$ to every of the $N_\mathrm{tet}$ individual tetrahedra of the mesh, as shown in \fref{fig-cell-model}(a).
It is labeled ``Random'' in the plot.

\subsection{Homogeneous equivalent cell model}
\label{sec-methods-equivalent-cell-model}
\begin{figure}
	\caption{\label{fig-cell-model}(a)~Definitions of our inhomogeneous cells and their homogeneous equivalent showing the stiffness ratio of the individual tetrahedra. (b)~Our inhomogeneous cell under compression at different values of the deformation $\deformation$. (c)~The stationary cell shape of a cell with a centered inhomogeneity in linear shear flow with increasing capillary number $\capillarynumber$. (d)~Our inhomogeneous cell flowing through a cylindrical capillary migrates towards the symmetry axes while maintaining an ellipsoidal shape. At the center, it assumes a bullet-like shape.
}
\includegraphics[width=\linewidth]{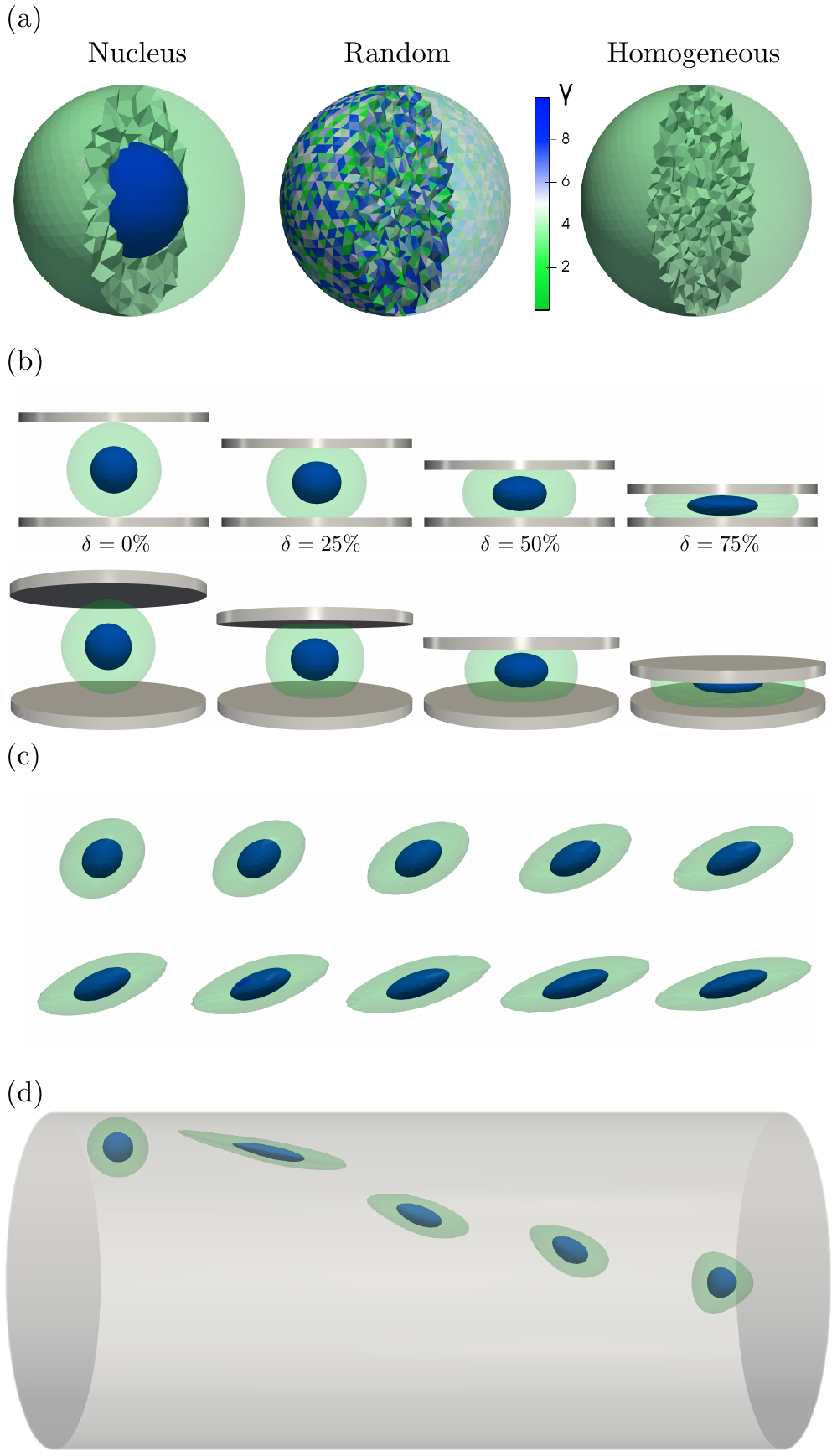}
\end{figure}

We construct a simplified but equivalent description of the inhomogeneous cell model from \sref{sec-methods-inhomogeneous-cell-model}, shown in \fref{fig-cell-model}(a).
It is labeled ``Homogeneous'' in the plot.
The same hyperelastic computations are performed on a tetrahedralized, initially spherical, mesh.
Instead of the spatially inhomogeneous stiffness distribution, however, a single parameter is computed by volume weighted averaging the constituents.
The effective Young's modulus of our equivalent cell model is defined as:
\begin{align}
\label{eq-effective-youngs-modulus}
	\youngseffective = \frac{1}{\volumecell}\qty( \volumeshell\youngsshell + \volumecore\youngscore ) = \qty[ 1 + \qty(\stiffnessratio-1) \sizeratio^3 ] \youngsshell
\end{align}
to substitute the inhomogeneous cell with nucleus.
Analogously for our random inhomogeneous cell model, the effective Young's modulus is computed as
\begin{align}
	\label{eq-effective-youngs-modulus-random}
	\youngseffective = \frac{1}{\volumecell} \sum\limits_{i=1}^{N_\mathrm{tet}} V_i \youngsshell \stiffnessratio_i 
	= \youngsshell \stiffnessratio_\mathrm{eff}\eqspace ,
\end{align}
from the volumes $V_i$ and the Young's moduli $E_i = \stiffnessratio_i \youngsshell$ of the $N_\mathrm{tet}$ individual tetrahedra.
In our setup, the volume averaged stiffness ratio is $\stiffnessratio_\mathrm{eff}\approx5.5$.

\subsection{Cell simulations under compression}
\label{sec-methods-afm-compression}
We create a compression scenario similar to mechanical characterization techniques of cells via atomic force microscopy by compressing our model between an upper, moving, and a lower, resting, plate using the algorithm from~\cite{muller_hyperelastic_2021}.
From our quasistatic simulation we obtain the normal force $\force$ exerted by the upper plate onto the cell, which causes a deformation as shown in \fref{fig-cell-model}(b).
We define the deformation parameter $\deformation$ as the relative compression, \ie, the plate-plate distance divided by the cell diameter.
We perform our simulations up to very large deformations of $\deformation=\SI{75}{\percent}$ for parameters $\stiffnessratio\in\qty{0,2,10,20}$ and $\sizeratio\in\qty{0.1,0.2,\dots,0.9}$, as well as for our random inhomogeneous cell.
We then perform another set of simulations with our homogeneous equivalent cell with the effective Young's modulus from \eqref{eq-effective-youngs-modulus} and \eqref{eq-effective-youngs-modulus-random}.

\subsection{Cell simulations in shear flow}
\label{sec-methods-shear-flow}
As a first flow scenario we use a linear shear flow, where our initially spherical deforms into an ellipsoidal body that undergoes a tank-treading motion.
To do so, we couple our hyperelastic tetrahedralized mesh to a Lattice Boltzmann flow simulation~\cite{kruger_lattice_2017,roehm_lattice_2012-1,limbach_espressoextensible_2006} via an immersed-boundary algorithm~\cite{schlenk_parallel_2018,bacher_clustering_2017}, using the same procedure as in \cite{muller_hyperelastic_2021,mueller_predicting_2022}.

Since the cell assumes an ellipsoidal shape, we choose the Taylor deformation parameter~\cite{muller_hyperelastic_2021,saadat_immersed-finite-element_2018}
\begin{align}
	\label{eq-def-taylor-deformation}
	\taylordeformation = \frac{a-b}{a+b}
\end{align}
with the ellipsoids major and minor semi-axis, respectively $a$ and $b$, as our measure for the cell deformation.
In analogy to the normal force introduced in \sref{sec-methods-afm-compression}, the strength of the shear flow is best characterized using the dimensionless shear rate, or capillary number
\begin{align}
	\label{eq-def-capillary-number}
	\capillarynumber =\frac{\dynvisc \shearrate}{\shearmodshell} = 2\qty(1+\poissonratio)\frac{\dynvisc \shearrate}{\youngsshell}\eqspace ,
\end{align}
where $\dynvisc$ denotes the surrounding fluid's dynamic viscosity and $\shearrate = \pdv{\velocity_x}{y}$ the constant velocity gradient.
Commonly, the shear modulus $\shearmodulus$ is used as stiffness parameter for this definition.
It relates to the Young's modulus of the previous section via the Poisson's as $\youngsmodulus=2\qty(1+\poissonratio)\shearmodulus$.
Hence, the stiffness ratios $\stiffnessratio$ and $\stiffnessratio\effindex$ have the identical value when defined analogously  to \eqref{eq-def-stiffness-ratio} and \eqref{eq-effective-youngs-modulus-random} via the shear moduli of the nucleus and the cytoskeleton, respectively $\shearmodcore$ and $\shearmodshell$.
\\
In \fref{fig-cell-model}(c), we show the stationary shape of our inhomogeneous cell at various $\capillarynumber$.
In addition to the ellipsoidal deformation of the entire shape, we find that the centered inhomogeneity, too, deforms into an ellipsoidal manner. 
However, its isolated deformation is visibly less pronounced.
We perform our simulations for $\stiffnessratio\in\qty{0,2,10,20}$ and $\offsetratio\in\qty{0,0.45}$, and with our random inhomogeneous cell.
Using the effective shear modulus $\shearmodeff$ in \eqref{eq-def-capillary-number}, we compare the inhomogeneous cells' behavior with the master curve describing the homogeneous equivalent cell.

\subsection{Cell simulations in capillary flow}
\label{sec-methods-pipe-flow}
In our second flow scenario, we place the initially spherical cell inside a cylindrical pipe with radius $\channelradius$, where an axial pressure gradient $G$ drives the Poiseuille flow~\cite{muller_flow_2020-1}.
Here, we need to distinguish two important cases, as illustrated in \fref{fig-cell-model}(d):
(i) When placed off-centered the cell will assume an approximately ellipsoidal shape according to the local shear rate.
Recently, it has been shown experimentally~\cite{gerum_viscoelastic_2022-2} and numerically~\cite{mueller_predicting_2022}, that a local shear flow approximation is valid for microfluidic and pipe flow applications, given that cells flow off-centered.
The local Capillary number as function of the radial position $r$ is given by
\begin{align}
	\label{eq-local-ca-pipeflow}
	\capillarynumber\qty(r) = \frac{G}{2\shearmodshell} \frac{r}{\channelradius}
\end{align}
Due to the fluid's shear stress, however, the cell continuously migrates from its starting point towards the center where the local shear flow approximation becomes insufficient.

(ii) At the channel axes the cell assumes a bullet-like shape due to the symmetrical flow conditions, as shown in \fref{fig-cell-model}(d).
This shape can be characterized by its strain in axial and radial direction, which we define as the maximum elongation in the respective direction divided by the cells reference diameter:
\begin{align}
	\strain_x  \defeq \frac{l_x}{2\cellradius}
	\quad \mathrm{and} \quad 
	\strain_r  \defeq \frac{l_r}{2\cellradius}
\end{align}
We perform our simulations for $\stiffnessratio\in\qty{0,2,10}$ and with the random inhomogeneous cell and compare the results to those of the homogeneous equivalent cell.

\section{Results}
\label{sec:results}
\subsection{Cells under compression}
\label{sec-results-compression}
\begin{figure}
	\caption{\label{fig-compression-1}The force versus deformation behavior of our proposed homogeneous equivalent cell compared to inhomogeneous cells with stiffer nucleus and our random inhomogeneous cell. Increasing the stiffness of the nucleus increases the overall force necessary to compress the cell. 
	}
	\includegraphics[width=\linewidth]{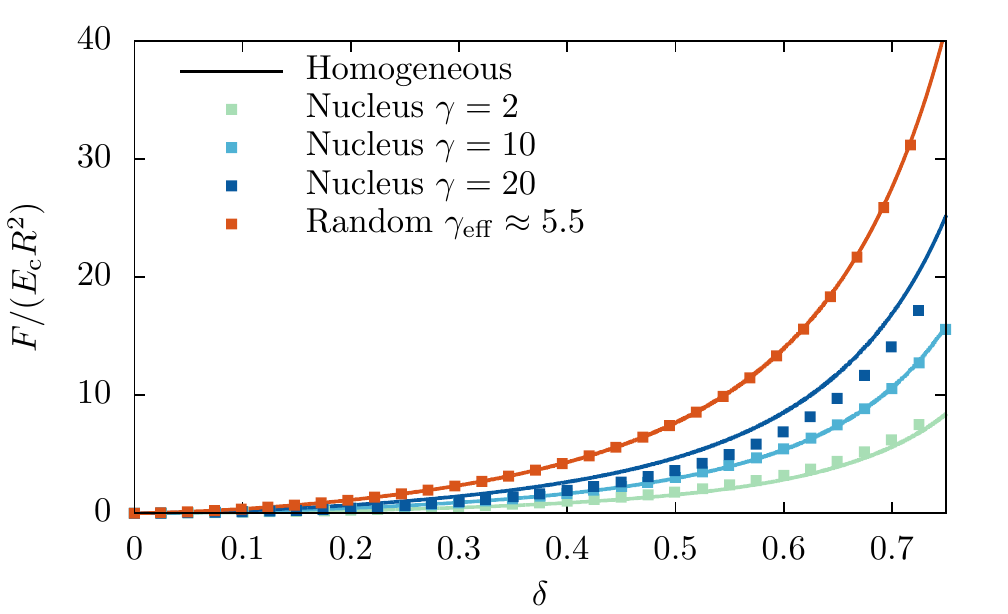}
\end{figure}

\begin{figure}
	\caption{\label{fig-compression-deviation}
	 (a)~The quality of the homogeneous equivalent can be visualized using the normalized force $\force^*$ \protect\eqref{eq-def-normalized-force}.
	 (b)~Variation of the volume ratio $\sizeratio$ \eqref{eq-def-stiffness-ratio}.
	The insets show a close-up view revealing that parameter variations of $\stiffnessratio$ and $\sizeratio$ do not affect the quality in a consistent manner.
}
\includegraphics[width=\linewidth]{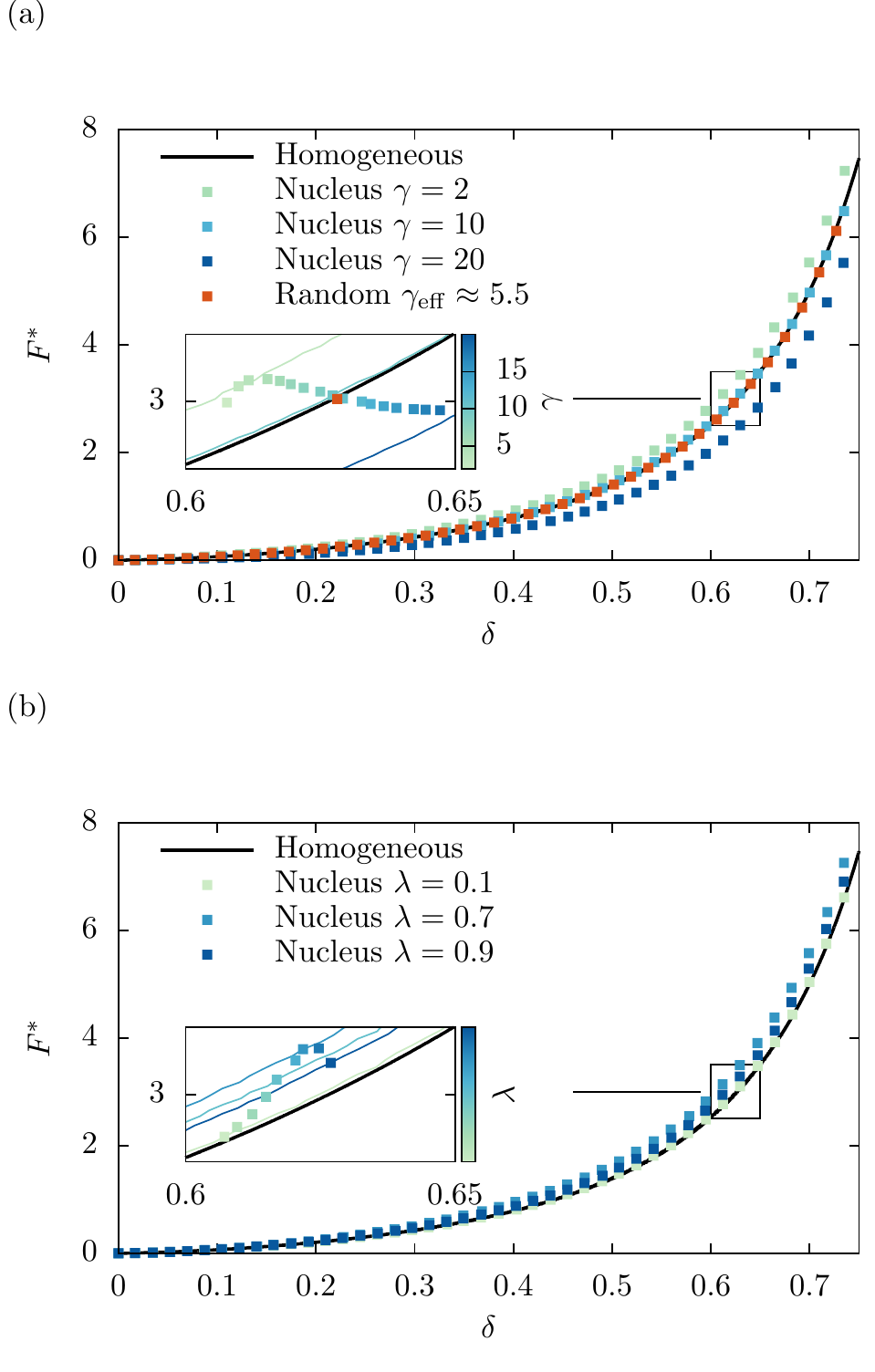}
\end{figure}

We first place a spherical nucleus with $\sizeratio=\half$ at the center of the cell and perform the compression simulations.
When increasing the stiffness ratio $\stiffnessratio$ at a constant size of the nucleus under compression, we find that --- as expected --- the force needed to compress the whole cell to a certain deformation $\deformation$ increases.
This is shown in \fref{fig-compression-1}, where we plot the dimensionless force $\force/(\youngsshell\cellradius^2)$ versus the deformation for our inhomogeneous cells with nucleus.
It is normalized using the Young's modulus of the shell $\youngsshell$ and hence identical for all simulations with different $\stiffnessratio$.
In the same manner, we plot in \fref{fig-compression-1} the data obtained from the simulations performed with the corresponding homogeneous equivalent cells as lines.
We find that, even for a nucleus $20$ times stiffer than the cytoskeleton, the deviation from the homogeneous equivalent cell are not significant.
Interestingly, our data for $\stiffnessratio=10$ matches perfectly with its homogeneous equivalent with $\youngseffective=2.125\youngsshell$, whereas the differences for other values of $\stiffnessratio$ deviate in different directions.
While for $1<\stiffnessratio<10$, the inhomogeneous cell exhibits stronger strain hardening than the homogeneous equivalent cell, for $\stiffnessratio>10$ the strain hardening is instead decreasing.
\\
This deviation can be visualized in a more quantitative way when the force is non-di\-men\-sion\-al\-ized using the corresponding effective Young's modulus \eqref{eq-effective-youngs-modulus}, giving
\begin{align}
	\label{eq-def-normalized-force}
 	\force^* = \frac{\force}{\youngseffective \cellradius^2} \eqspace ,
\end{align}
which is shown in \fref{fig-compression-deviation}(a).
Due to this non-di\-men\-sion\-al\-iz\-a\-tion, all data curves describing homogeneous cells collapse onto one master curve.
The data of the inhomogeneous cells then deviates from this master curve in different directions, which indicates the quality of the homogeneous equivalent description.
It is apparent that the variation of $\stiffnessratio$ does not lead to a consistent deviation from the homogeneous description, but instead changes in different directions.
This is visualized in the inset of \fref{fig-compression-deviation}(a), where data for additional values of $\stiffnessratio$ is plotted.

We now vary the size of the inhomogeneity at constant stiffness $\stiffnessratio=2$ between the two limiting cases that describe homogeneous cells, namely $\sizeratio=0$ (all softer shell) and $\sizeratio=1$.
In \fref{fig-compression-deviation}(b), the resulting normalized force \eqref{eq-def-normalized-force} versus deformation curves show little deviation, and they always lie between the curves for $\stiffnessratio=1$ and $\stiffnessratio=2$ from \fref{fig-compression-1}.
The inset of \fref{fig-compression-deviation}(b) shows the increase and decrease of the deviation from the homogeneous cases, with a maximum value at around $\sizeratio \approx 0.7$.
Note that this value is close to, yet differs, from the value $\sizeratio=2^{-\frac{1}{3}}\approx0.79$ obtained for equal volumes of shell and inhomogeneity.

\begin{figure}
	\caption{\label{fig-compression-ellipsoidal-inhomogeneity}(a)~Variation of the position of the nucleus along two independent axes slightly increases the accuracy of the homogeneous description. (b)~A nucleus with ellipsoidal shape (but same volume) comes without notable effect on the force versus deformation behavior independent of the orientation.
	}
\includegraphics[width=\linewidth]{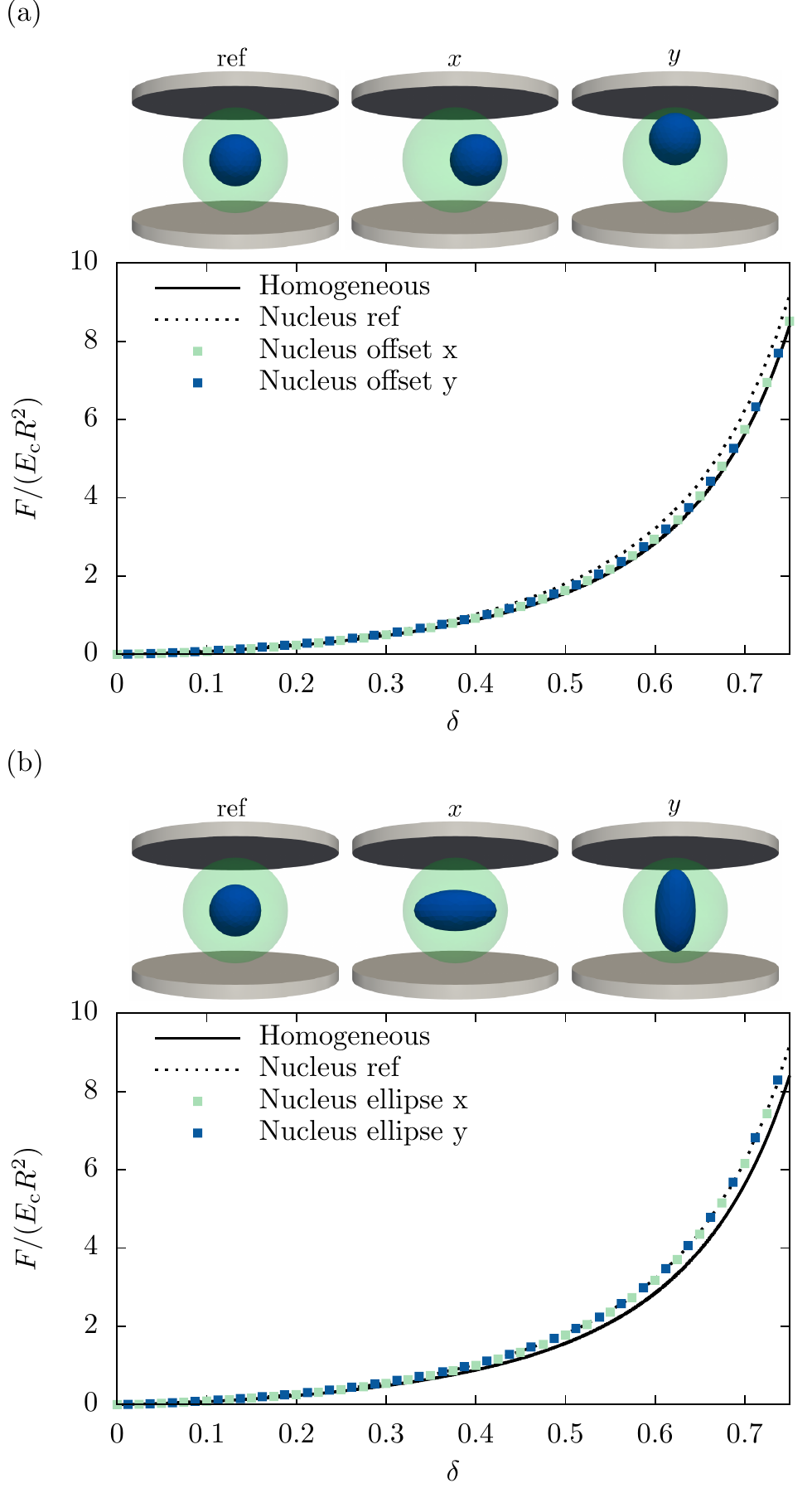}
\end{figure}

Next, we move the inhomogeneity ($\stiffnessratio=2$ and $\sizeratio=\half$) away from the center and very close to the cell surface, \ie, $\offsetratio=0.45$.
As illustrated in \fref{fig-compression-ellipsoidal-inhomogeneity}(a), we denote with $x$ the direction parallel to the plates and with $y$ the perpendicular direction.
We deduce the insignificance of the position of the inhomogeneity from the force versus deformation curves in \fref{fig-compression-ellipsoidal-inhomogeneity}(a), where all data points overlap exactly.

Finally, we alter the shape of the nucleus and replace the centered spherical inclusion with an ellipsoid of equal volume with semi-axes $a\approx 0.8\,\cellradius$, $b=c\approx 0.4\,\cellradius$.
We choose again the parallel ($x$) and perpendicular ($y$) alignment of the major semi-axis, which we compare to the centered spherical inclusion ($\sizeratio=\half$) denoted with $\mathrm{ref}$, as shown in \fref{fig-compression-ellipsoidal-inhomogeneity}(b).
The resulting force versus deformation curves for ($\stiffnessratio=2$) in \fref{fig-compression-ellipsoidal-inhomogeneity}(b) underline that a variation of the inhomogeneity's shape effectively does not affect the compression behavior of a cell. Due to imperfections in the mesh and our simulation method, cells with their ellipsoidal nucleus aligned along the y-axis tend to rotate during deformation when using a greater stiffness ratio (broken rotational symmetry, non zero torque). We therefore discuss only the $\stiffnessratio=2$ case, where there is no rotation present and the results of both geometries are accurate.

We then perform the same simulation with our random inhomogeneous cell model from \fref{fig-cell-model}(a).
The force versus deformation behavior in \fref{fig-compression-1} and \fref{fig-compression-deviation}(a) excellently matches with its homogeneous equivalent cell.

This section shows that, for compression scenarios, a heterogeneous cell can in practice be replaced with a homogeneous equivalent cell with a volume averaged Young's modulus, since neither the stiffness difference nor the size, the position, or the shape, of the inhomogeneity have a significant impact on the force necessary to produce a certain cell deformation.

\subsection{Cells in linear shear flow}
\label{sec-results-shear}
\begin{figure}
	\caption{\label{fig-shearflow}
		~Taylor deformation of the nucleate cell for different capillary number in shear flow. The dotted lines represent the curves of their respective homogeneous equivalent cells. $\stiffnessratio$.
	}
\includegraphics[width=\linewidth]{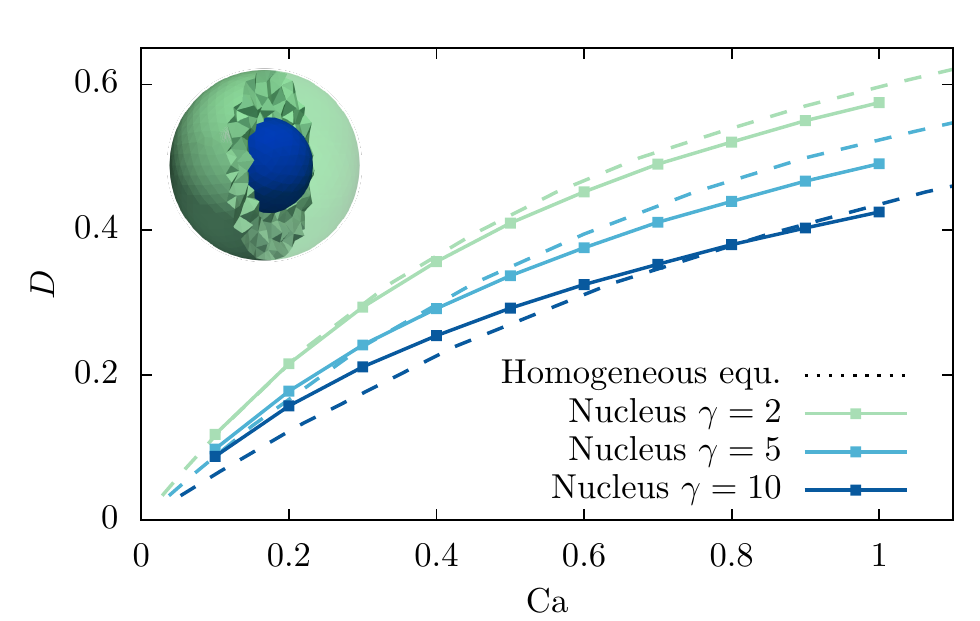}
\end{figure}

For our investigations of cells in flow, we start by putting our initially spherical inhomogeneous cell ($\stiffnessratio=2$, $\sizeratio=\half$, and $\offsetratio=0$) in a linear shear flow with shear rate $\shearrate$.
After a transient time span its shape becomes stationary (cf.~\fref{fig-cell-model}(c)), and the cell undergoes a continuous tank-treading motion.
We find in \fref{fig-shearflow} excellent agreement between our nucleate cell and its homogeneous equivalent cell, when plotting the stationary value of $\taylordeformation$ obtained from the simulation with $\stiffnessratio=2$ as a function of the Capillary number \eqref{eq-def-capillary-number}.
In accordance with the compression simulations of \fref{fig-compression-1}(a), we find that the inhomogeneous cell with nucleus at a stiffness ratio $\stiffnessratio=2$ yields a slightly lower deformation than its homogeneous equivalent.
Similarly in \fref{fig-shearflow}, we depict additionally the results for $\stiffnessratio=5$ and $10$.
When comparing the inhomogeneous cell to its respective homogeneous equivalent we find good agreement over a wide range of deformations. 
Only at low deformations ($<0.02$) the equivalent cell model yields inaccurate results, which is to be expected from previous studies~\cite{cao_evaluating_2013}.
In that range, the data approaches the results as they would be obtained from a homogeneous cell with $\shearmodshell$ throughout.
Analogously to our observation in the compression setup in \fref{fig-compression-1}(a), the data for $\stiffnessratio=10$ is surprisingly accurate for large deformation.

\begin{figure}
	\caption{\label{fig-shearflow-offset}
		(a)~Time series snapshots for the inhomogeneous cell with nucleus displaced parallel to the shear plane. Due to the rotation of the cellular material, the bump produced by the inhomogeneity travels around the cell.
		(b)~Time development of the Taylor deformation parameter $\taylordeformation$ for an inhomogeneous cell with the nucleus displaced parallel to the shear plane oscillating around that for a displacement perpendicular to the shear plane.
		(c)~Average $\taylordeformation$ of the cell with off-centered nucleus compared to the data of \fref{fig-shearflow}(a).}
\includegraphics[width=\linewidth]{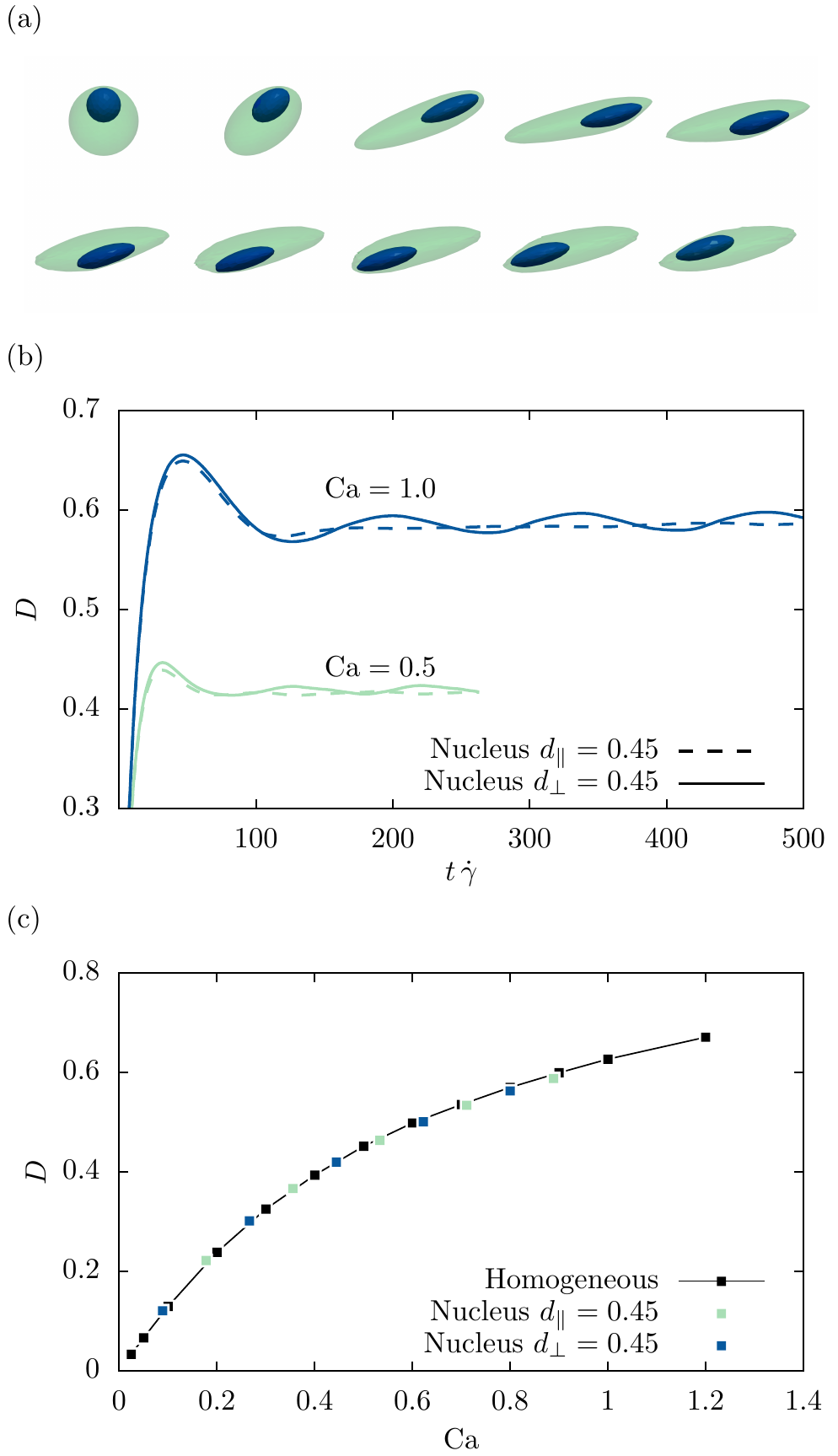}
\end{figure}

A significant influence on the dynamic behavior is found when the stiffer nucleus is not centered in the cell.
Our cell with nucleus at $\offsetratio_\parallel=0.45$ is show in \fref{fig-shearflow-offset}(a).
A series of snapshots depicts the tank-treading motion of the entire cell, where the nucleus produces a bump at the cell surface, which periodically moves along.
This behavior is also reflected in the time development of the Taylor deformation \eqref{eq-def-taylor-deformation}, as shown in \fref{fig-shearflow-offset}(b) (dashed lines), where $\taylordeformation$ is plotted as function of the dimensionless time $t\shearrate$.
If instead the nucleus in placed perpendicular to the shear plane at $\offsetratio_\perp=0.45$, the same stationary behavior as for a centered nucleus is obtained (solid lines).
The time average of the deformation in this state is shown in \fref{fig-shearflow-offset}(c) for both nucleus offsets as well as the corresponding homogeneous equivalent cell.
It becomes clear that the time-averaged deformation of the cell with off-centered nucleus is perfectly covered by our homogeneous description.


\subsection{Cell in capillary flow}
\label{sec-results-pipe}

\begin{figure}
	\caption{\label{fig-pipeflow-centered-deformation}(a)~Snapshots of the inhomogeneous cell with nucleus and its homogeneous equivalent when flowing at the center of the pipe for $\stiffnessratio=2$ and $10$. The gray area shows a slice of the homogeneous equivalent cell while the green/blue overlay depicts the inhomogeneous cell with nucleus. (b)~The stationary axial ($\strain_x < 1$) and radial strains ($\strain_r > 1$) of our inhomogeneous cells as function of the stiffness ratio $\stiffnessratio$ in comparison to those of the homogeneous equivalent cell. Error bars denote the standard deviation.}
\includegraphics[width=\linewidth]{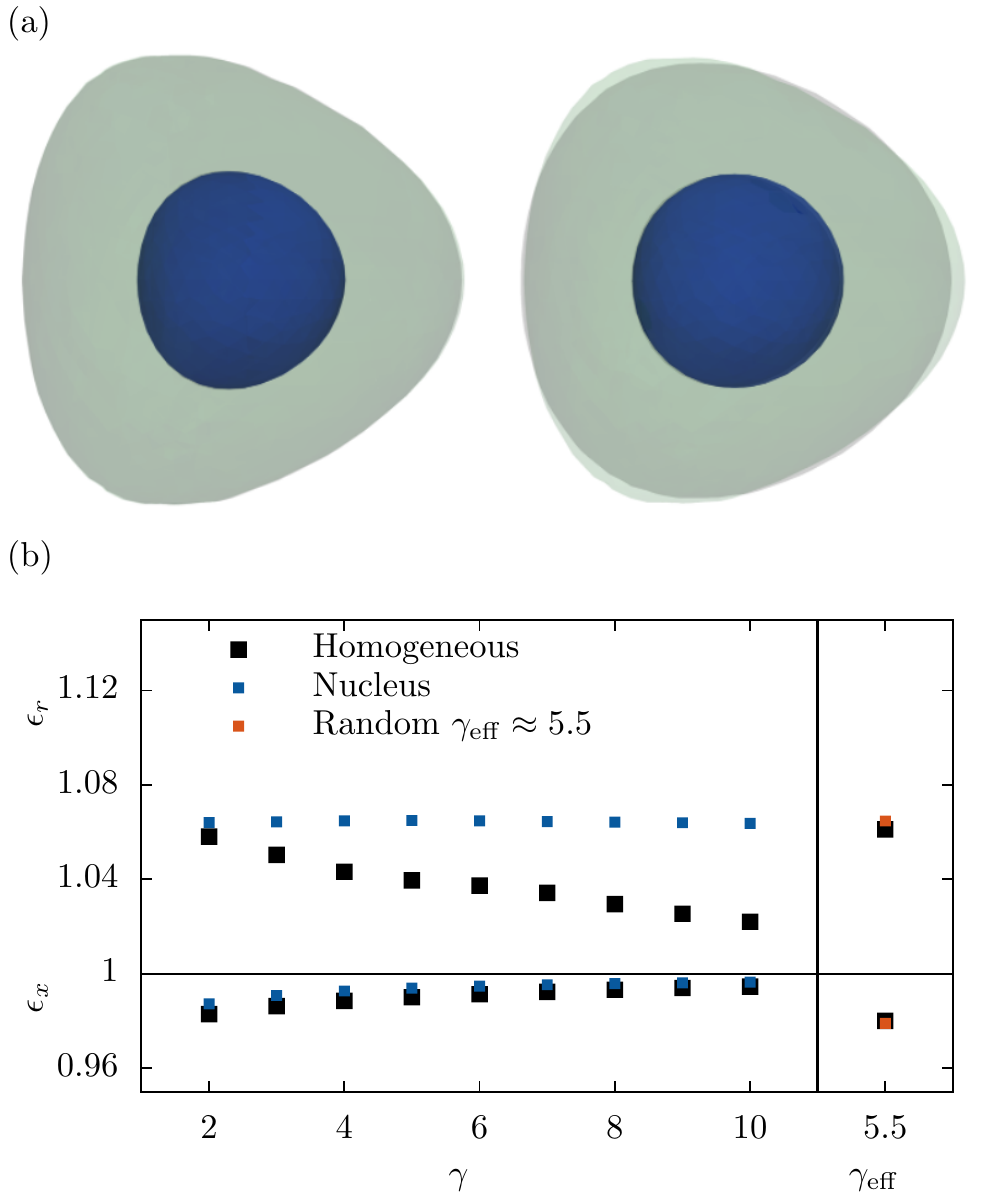}
\end{figure}

The two major differences between the pressure driven flow through a pipe or microchannel and the simple shear flow scenario are (i)~the non-linearity of the velocity profile and (ii)~ the symmetry conditions at the channel axis.
When the cell flows at the center of the channel, it assumes a stationary bullet-like shape as depicted in \fref{fig-pipeflow-centered-deformation}(a) for our inhomogeneous cell with $\stiffnessratio=2$ and $10$, as overlay over its homogeneous equivalent.
Visible, these two shapes agree perfectly with each other, even though the cytoskeleton in contact with the surrounding fluid is $\num{2.125}$ times softer for the inhomogeneous cell.
As for a quantitative analysis, the axial and radial strains, $\strain_x$ and $\strain_r$, assume a stationary value after a short time span.
We depict the time average of these stationary values for our inhomogeneous cell with nucleus, our random inhomogeneous cell, and their respective homogeneous equivalent in \fref{fig-pipeflow-centered-deformation}(b).
It becomes apparent that the  radial strain of nucleate cell shows an increasing deviation from its homogeneous equivalent, while the axial strain remains accurate.
In contrast to the decreasing radial strain of the homogeneous equivalent cell, which is simply explained by its larger stiffness, the radial strain of the nucleate cell remains almost unchanged.
This can be understood the following way:
On the one hand, the soft cytoskeleton of the nucleate cells has the same stiffness throughout all simulations.
Since all simulations are performed using the same flow conditions, a similar stress is acting on the cell surface and the cytoskeleton.
The stiffer nucleus, on the other hand, is centered inside the cell and located on the symmetry axis of the channel, where the fluid stress vanishes~\cite{muller_flow_2020-1}.
We can therefore assume a weaker influence of the nucleus in this scenario as compared to the off-centered flow, in which the nucleus itself was subjected to large stresses.
This observation is underlined by our investigation of the random inhomogeneous cell, which shows excellent agreement with its homogeneous equivalent in  \fref{fig-pipeflow-centered-deformation}(b).
Thus, we conclude that our proposed homogeneous equivalent description is still valid in capillary flow.

\section{Conclusion}
In this work, we presented systematic numerical demonstration that the mechanics of highly inhomogeneous cells can be replaced with a simple homogeneous equivalent  by means of a straightforward volume averaged effective elasticity.
\\
For this, we constructed three numerical cell models: a homogeneous cell, a cell including a well-defined inhomogeneity, and a random inhomogeneous cell.
All models showed the same force versus imposed deformation behavior under AFM-like compression.
In shear and pipe flow simulations, we found that an inhomogeneity can have an impact on the dynamic time evolution of the cell's shape.
However, no difference in the stationary behavior was observed and the average strain as function of the fluid forces agrees exactly.
\\
Our proposed homogeneous equivalent hence stays valid under different loading scenarios and is independent of the shape, size, stiffness, or distribution, of the cell's internal heterogeneity.
Our results thus validate in hindsight the simplifying approaches taken in many previous experimental and computational works, but also provide a solid basis on which future experimental data can be analyzed and phyisically reliable computer simulations can be constructed.

\section{Appendix 1: Mooney-Rivlin strain energy computations}

\begin{figure}
	\caption{\label{fig-Mooney-Rivlin} The force versus deformation behavior of inhomogeneous cells using the Mooney-Rivlin model. The cell parts material constants are set fractions of the given shear modulus: $\mu_1 = 0.25\mu, \mu_2 = 0.75\mu$. The homogeneous equivalent cell approximation is as good as when using the Neo-Hookean model.}
	\includegraphics[width=\linewidth]{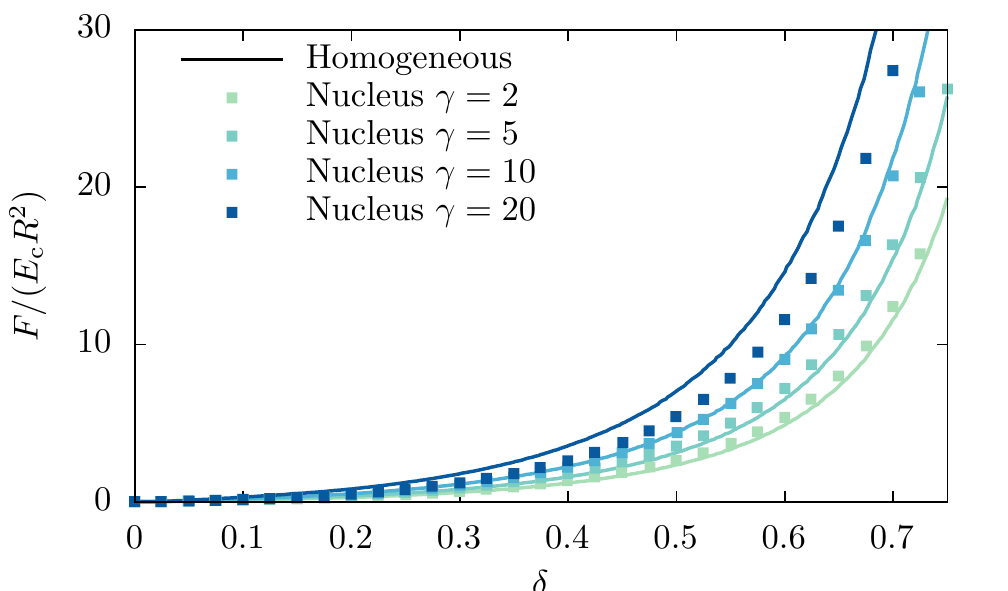}
\end{figure}

As described in section \ref{sec-methods-nucleus-cell-model} we use Neo-Hookean strain energy computations for the cells dynamics. The more sophisticated Mooney-Rivlin model in \cite{muller_hyperelastic_2021} uses two separate material constants $\mu_1$ and $\mu_2$ for computing the strain energy density of a tetrahedron.
We tested whether the effective Youngs modulus model for a homogeneous equivalent cell still works when using the Mooney-Rivlin model by deriving the shear modulus: $\mu = \mu_1 + \mu_2$ . Figure \ref{fig-Mooney-Rivlin} shows that this is the case which further supports the validity of our model.

\section*{Acknowledgements}
	Funded by the Deutsche Forschungsgemeinschaft (DFG, German Research Foundation) --- Project number 326998133 --- TRR 225 ``Biofabrication'' (subprojects B07). 
	We further acknowledge support through the computational resources provided by the Bavarian Polymer Institute.

\renewcommand{\emph}[1]{#1}
\bibliography{literature.bib}   
\renewcommand{\emph}[1]{\underline{#1}}
%

\end{document}